\documentclass{osa-article}

\journal{oe}
\usepackage{lineno}

\usepackage{amsmath}
\usepackage{notes2bib}

\usepackage[utf8]{inputenc} 
\usepackage{graphicx, xcolor}   
\usepackage{physics}
\usepackage[normalem]{ulem}
\usepackage{bm}

\newcommand{\onlinecite}[1]{\hspace{-1 ex} [\nocite{#1}\citenum{#1}]} 

\newcommand{\ex}{\mathbf{e}_x}
\newcommand{\ey}{\mathbf{e}_y}
\newcommand{\ez}{\mathbf{e}_z}

\newcommand{\Isat}{I_{\rm sat}}

\DeclareMathAlphabet{\mathpzc}{OT1}{pzc}{m}{it}

\def\Rb87{^{87}\mathrm{Rb}}                             
\def\K40{^{40}\mathrm{K}}                    		    

\begin{document}

\title{Direct Calibration of Laser Intensity via Ramsey Interferometry for Cold Atom Imaging}

\author{Emine~Altunta\c{s}\authormark{1,*} and I.~B.~Spielman\authormark{1,2}}
\address{\authormark{1}Joint Quantum Institute, National Institute of Standards and Technology, and University of Maryland, Gaithersburg, Maryland, 20899, USA\\
	\authormark{2}ian.spielman@nist.gov \\
	\email{\authormark{*}altuntas@umd.edu}
}
\date{\today}

\begin{abstract*}
A majority of ultracold atom experiments utilize resonant absorption imaging techniques to obtain the atomic density.
To make well-controlled quantitative measurements, the optical intensity of the probe beam must be precisely calibrated in units of the atomic saturation intensity $\Isat$.
In quantum gas experiments, the atomic sample is enclosed in an ultra-high vacuum system that introduces loss and limits optical access; this precludes a direct determination of the intensity.
Here, we use quantum coherence to create a robust  technique for measuring the probe beam intensity in units of $\Isat$ via Ramsey interferometry.
Our technique characterizes the ac Stark shift of the atomic levels due to an off-resonant probe beam.
Furthermore, this technique gives access to the spatial variation of the probe intensity at the location of the atomic cloud.
By directly measuring the probe intensity just before the imaging sensor our method in addition yields a direct calibration of imaging system losses as well as the quantum efficiency of the sensor. 
\end{abstract*}

\section{Introduction} \label{sec:Introduction}

Ultracold atom experiments are now entering an era of precision quantum simulation; this poses new challenges to the underlying measurement methodology.
These systems are most commonly measured with on- or near-resonant laser light; as such calibrating the optical intensity can be important for obtaining accurate measurements.
In particular for sufficiently intense light the atomic transition becomes saturated and each atom achieves a maximum scattering rate.
For example, absorption imaging illuminates an atomic ensemble with a resonant probe laser and measures the atoms' shadow.
In order to quantify the number of atoms from the absorbed light the probe intensity in units of the saturation intensity $\Isat$ must be known.
Because ultracold atoms exist in an ultrahigh vacuum  environment, it is difficult to calibrate the laser intensity at their location, and losses from vacuum viewports make {\it ex situ} measurements inaccurate.
This paper describes a robust technique for determining the probe light intensity for ultracold atom measurements using Ramsey interferometry (RI).

For resonant imaging techniques, the signal-to-noise ratio (SNR) typically reaches its maximum with intensities $I \approx \Isat$.
With modern low noise, high efficiency detectors, photon shot noise is generally the leading noise source.
As such, for $I \ll \Isat$ the scattering rate is proportional to $I$ and the SNR scales like $I^{1/2}$.
By contrast for $I \gg \Isat$ the scattering rate is independent of $I$ and the SNR scales like $I^{-1/2}$ (see App.~\ref{App:SNR} for more detail).
When $I\ll\Isat$ the numerical value of the probe intensity $I$ drops out of the expressions for the atomic density, however, it contributes significantly when $I \gtrsim \Isat$.

Several techniques have been developed to overcome the difficulty of measuring the probe laser {\it in vacuo}.
The most straightforward technique~\cite{Reinaudi2007} relies on simply measuring the reduction in absorption for increasing intensity to estimate $\Isat$.
This quick, easy-to-implement method serves to produce a measure of atomic density that is independent of probe intensity.
However, as a heuristic modeling approach there is no assurance of an accurate determination of $\Isat$.
A more recently developed method overcame this modeling limitation by using the acceleration from radiation pressure to directly count the average number of scattered photons~\cite{Hueck2017}.
Although this is a more direct technique, it adds technical complexity.
Here, we describe a technically simple approach that yields a spatially resolved map of the intensity across the atomic ensemble.

\begin{figure}[tb!]
\begin{center}
\includegraphics{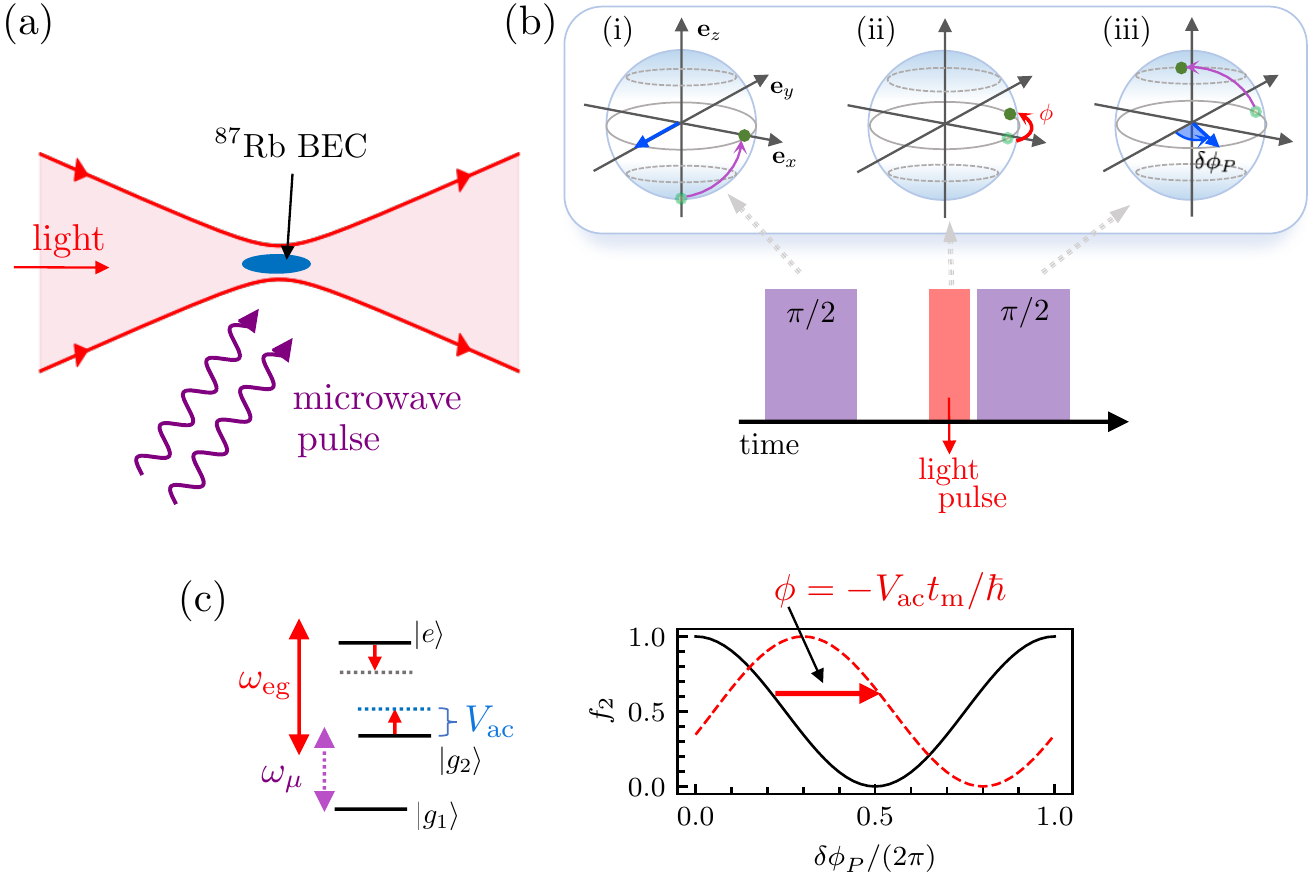}
\end{center}
\caption[Measurement principle]{Principle of measurement.
(a) A microwave pulse (wiggly purple curves) transforms the system (blue) into a superposition state and subsequently the BEC is briefly illuminated with far-detuned laser light (red).
(b) Top: Bloch sphere representation of a Ramsey interferometry sequence, where the dark blue arrows indicate the axes of rotation for each microwave pulse; the purple arrows depict the associated trajectories; and the green circles mark the state before (translucent) and after (solid) each step . 
The red arrow shows the evolution from the ac Stark shift associated with the laser pulse.
Bottom: Underlying time sequence.
(c) Left: Simplified level diagram.
The resonant microwave field (purple) with frequency $\omega_\mu$, couples $\ket{g_1}$ and $\ket{g_2}$.
The off-resonant laser light (red) of frequency $\omega_{\rm{ge}}$ couples $\ket{g_2}$ to $\ket{e}$ and induces an ac Stark shift (blue) on $\ket{g_2}$. 
Right: Idealized Ramsey signal without (black) and with (dashed red) the off-resonant laser light, showing the phase shift resulting from the ac Stark shift.
}
\label{Fig1:Principle}
\end{figure}

Rather than focusing on changes in absorbed light or radiation pressure, we consider the ac Stark shift, derived from the dispersive part of the light-matter interaction.
The ground state of a two level atom undergoes an ac Stark shift~\cite{Steck_QuantOpt} of 
\begin{align}
\frac{V_{\rm ac}}{\hbar} &= \frac{\Omega^2}{4\delta}
\label{Eqn:AcStark_ground}
\end{align}
from a laser beam with resonant Rabi frequency $\Omega$ detuned $\delta$ from resonance; here $\hbar$ is the reduced Planck's constant.
The saturation intensity $\Isat$ is a useful parameter that relates the
intensity $I$ to the natural linewidth $\Gamma$ and the Rabi frequency via ${I}/{I_{\rm sat}} = 2\left|{ \Omega}/{ \Gamma}\right|^2$~\cite{Steck_QuantOpt}. 
Combining these expressions, we connect the ac Stark shift to the intensity with
\begin{align}
\frac{V_{\rm ac}}{\hbar \Gamma} &= \frac{1}{8} \frac{\Gamma}{\delta} \frac{I}{\Isat}.
\label{Eqn:VAcStark}
\end{align}
Our technique uses Ramsey interferometry (RI) to measure the  phase shift
\begin{equation}
\phi = -V_{\rm ac} t_{\rm m} / \hbar
\label{Eqn:RI_phase}
\end{equation} 
that an atomic ground state acquires in a time $t_{\rm m}$. 
Because $\delta$ and $t_{\rm m}$ are well known, our method yields a primary calibration of the laser intensity in units of $\Isat$.

This paper begins by formulating the basic principle of our measurement method via Ramsey interferometry in Sec.~\ref{sec:Method}.
Next, in Sec.~\ref{sec:ExpSys}, we continue with a more detailed description of our experimental setup and elaborate on the experimental implementation of our technique. 
In Sec.~\ref{sec:RamseyLightShift}, we present our Ramsey interferometry measurements and extract the spatially averaged laser intensity experienced by the atoms.  
Following, in Sec.~\ref{sec:PixByPix}, we utilize our Ramsey interferometry method to spatially characterize the intensity inhomogeneity in the region of the atomic system.
In Sec.~\ref{sec:IntensityVary}, we calibrate the the sensor using photo-electron shot noise; in conjunction with the $\Isat$ calibration this directly gives the full-system optical loss.


\section{Technique} \label{sec:Method}

Our experimental characterization of the ac Stark shift can be understood in terms of the three-level configuration shown in Fig.~\ref{Fig1:Principle}(c), consisting of ground states $\ket{g_1}$ and $\ket{g_2}$ (with energies $E_{g_1}$ and $E_{g_2}$) and excited state $\ket{e}$ (with energy $E_e$); the probe laser couples $\ket{g_2}$ to $\ket{e}$.
In this configuration $\ket{g_1}$ experiences no Stark shift\bibnote{In experimental reality $\ket{g_1}$ experiences a small Stark shift that we include as described in Sec.~\ref{sec:RamseyLightShift}.} and we use RI to measure the phase shift between the two ground states.

RI measures differential phase shift in a way analogous to a Mach–Zehnder interferometer, and operates by creating an interference between amplitude in a path that experiences a desired phase shift (here $\ket{g_2}$) and one that does not (here $\ket{g_1}$).
In practice we achieve this with atoms beginning in $\ket{g_1}$ and use an oscillatory microwave field with frequency $\omega_\mu = (E_{g_1} - E_{g_2})/\hbar$ [purple line in Fig.~\ref{Fig1:Principle}(a)] to couple to $\ket{g_2}$ with Rabi frequency $\Omega_R$. 
In this way, a $\pi/2$ pulse creates a superposition state $(\ket{g_1} + \ket{g_2}) / \sqrt{2}$ [Fig.~\ref{Fig1:Principle}(b)-i], and after a free evolution time $T$ [Fig.~\ref{Fig1:Principle}(b)-ii], a second $\pi/2$ pulse interferes the two paths [Fig.~\ref{Fig1:Principle}(b)-iii].
The final population in states $\ket{g_1}$ and $\ket{g_2}$ gives access to the differential phase shift acquired during free evolution.

This process is described by a sequence of evolution operators
\begin{equation}
\ket{\psi_{\rm f}} = \hat{R}(\theta)\hat{U} \hat{R}(0)\ket{g_1},
\label{Eqn:RamseyInt_AC}
\end{equation}
in which $\hat{U} = \ket{g_2}\bra{g_2} e^{i \phi} + \ket{g_1}\bra{g_1}$ describes the free evolution period including the phase shift $\phi$ from the probe pulse, and $\hat{R}(\theta) = \left[\hat I +i \left(\sin\theta \ \hat\sigma_x - \cos\theta\ \hat\sigma_y\right)\right]/\sqrt{2}$ describes a $\pi/2$ rotation about an axis residing in the $\ex$-$\ey$ plane of the Bloch sphere, rotated by an angle $\theta$ from $\ey$. 
In experiment, we select this angle by phase-shifting the second $\pi/2$ microwave pulse by $\theta = \delta\phi_P$ with respect to the first.
Here, $\hat{\sigma}_{x,y,z}$ are the Pauli operators in the ground state manifold and $\hat I$ is the identity.
Following this sequence, the occupation probabilities $f_{1,2}$ in $\ket{g_{1,2}}$ are measured.

We consider the simple case when $\hat U = \hat I$, i.e., no ac Stark shift, to elucidate the operation of this process.
In this case, the probability $f_2 = \left(1 + \cos \delta\phi_P\right)/2$, shown in black in Fig.~\ref{Fig1:Principle}(c), is an oscillatory function of the phase shift $\delta\phi_P$.
As shown by the red-dashed curve, any additional phase shift $\phi$ acquired during the free-evolution time gives $f_2 = \left[1 +  \cos (\delta\phi_P-\phi)\right]/2$.

Experimentally, we fit such data to 
\begin{equation}
f_2^{\rm expt} = \frac{1 + A\cos(\delta\phi_P - \phi)}{2} + b,
\label{Eqn:RI_fringe_fit}
\end{equation}
and obtain the phase shift $\phi$ as well as the contrast $A$, and center shift $b$.
Contrast reduction can result from experimental imperfections as well as spontaneous scattering, as described in terms of quantum back-action from a measurement perspective in Ref.~\onlinecite{Altuntas2022_RIL}.
The center shift results from differential losses between $\ket{g_1}$ and $\ket{g_2}$; averaged across our whole data set we find $b = -0.023(25)$, making it a small effect\bibnote{All uncertainties herein reflect the uncorrelated combination of single-sigma statistical and systematic uncertainties.}.
We then obtain the ac Stark shift $V_{\rm ac}$ with high accuracy by measuring the phase shift $\phi$ as a function of probe intensity $I$, detuning $\delta$, and pulse duration $t_m$.

\section{Experimental system} \label{sec:ExpSys}

\begin{figure}[tb!]
\begin{center}
\includegraphics[width=5.2in]{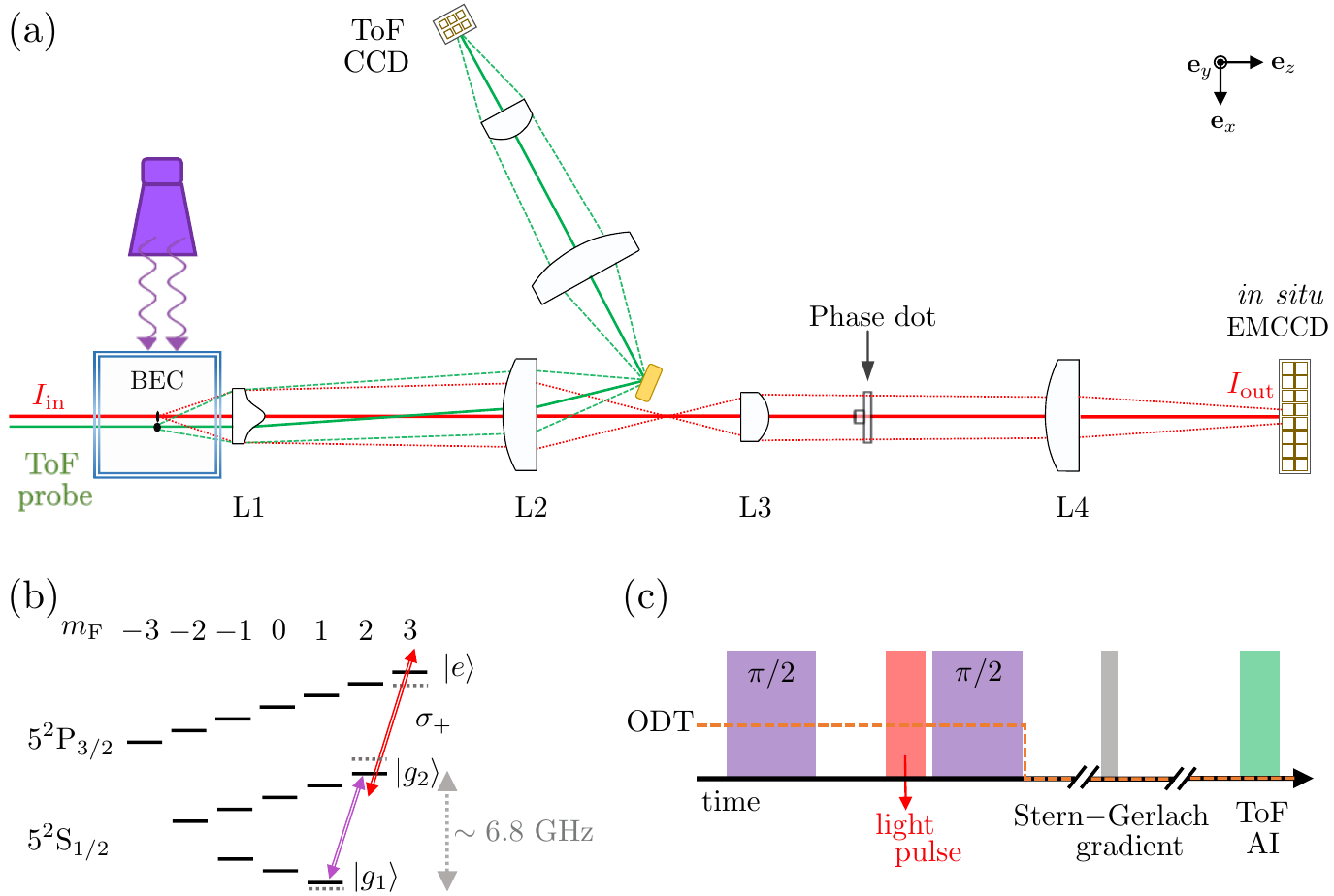}
\end{center}
\caption[Experimental imaging system]{Realistic imaging system, level diagram, and experimental sequence.
(a) Imaging setup. 
Our atomic system is imaged both {\it in situ} (with the main imaging system) and after TOF (with the auxiliary setup).
The off-resonant probe beam (red) illuminates the BEC {\it in situ} and the auxiliary imaging system uses a secondary laser beam (green). 
(b) Relevant hyperfine structure.
The purple arrow designates the microwave coupling; the red arrow designates the probe beam coupling on the $\rm{D}_2$ cycling transition of $\Rb87$.
The dashed lines indicate the effect of the ac Stark shift.
(c) RI time sequence.
An initial $\pi/2$ microwave pulse (purple) is followed by a $15\ \mu{\rm s}$ evolution period; then the off-resonant light pulse (red) is applied, and after a $5\ \mu{\rm s}$ delay, a second $\pi/2$ pulse (purple) completes the RI sequence.
The ODT (orange dashed) is turned off immediately following the Ramsey sequence.
A Stern-Gerlach gradient (grey) is applied during TOF, and the atoms are detected using standard absorption imaging (AI, green).
}
\label{Fig2:ExpSystem}
\end{figure}

Figure~\ref{Fig2:ExpSystem}(a) illustrates the fundamental challenge of calibrating the probe intensity at the location of atomic ensemble.
The red lines indicate our primary imaging path, with numerous optics separating the in-vacuum atoms from the sensor.
Optical losses in the imaging system render light intensity measurements at the imaging sensor inaccurate, while limited access at the location of the atoms due to the ultra-high vacuum system prevents direct in-vacuo measurements.
Here, we describe the experimental procedure used to acquire the ac Stark shift in RI measurements.

Our experiments commenced with highly elongated Bose-Einstein condensates (BECs) of $N = 0.70(15)\times10^5$ atoms confined in a $50\ {\rm mm}\times50\ {\rm mm}$ glass cell.
The BEC was trapped in a crossed optical dipole trap (ODT) realized by two 1064 nm beams propagating along $\ex$ and $\ey$ yielding trap frequencies $(\omega_x, \omega_y, \omega_z) = 2\pi \times \left[9.61(3), 113.9(3), 163.2(3) \right]\ {\rm Hz}$.
Trapped atoms were prepared in the $5^2\rm{S}_{1/2}$ electronic ground state in $\ket{g_1}\equiv\left|F = 1, m_{F} = 1\right\rangle$ as shown in Fig.~\ref{Fig2:ExpSystem}(b).
We implemented the RI $\pi/2$ pulses using a microwave magnetic field tuned to the $\approx 6.8\ {\rm GHz}$ ground state hyperfine splitting of $\Rb87$; our Rabi frequency was $\approx 7.5\ {\rm kHz}$. 
This formed the basis of our {\textit {in situ}} RI probe intensity calibration technique. 

Figure~\ref{Fig2:ExpSystem}(c) shows our overall time sequence for RI.
The first $\pi/2$ microwave pulse drive transitions between $\ket{g_1}$ and $\ket{g_2}\equiv\ket{F=2,m_F=2}$, creating the desired superposition state along the x-axis of the Bloch sphere. 
Then, our far-detuned probe laser of wavelength $\lambda\approx780\ {\rm nm}$ couples $\ket{g_2}$ to the $\ket{e}\equiv\ket{F'=3,m'_F=3}$ electronic excited state on the cycling transition, and weakly couples $\ket{g_1}$ to $\ket{e_2}\equiv\ket{F'=2,m'_F=2}$ (giving $\Isat\approx 1.67~\rm{mW/cm^2}$ for $\sigma_+$ polarization).
After brief $5\ \mu\rm{s}$ delay we applied the second $\pi/2$ microwave pulse with the $\delta\phi_P$ phase shift, closing the RI.
Subsequently, we extinguished the ODT to initiate a $20\ {\rm ms}$ time of flight (TOF).
During the TOF period, we spatially separated the hyperfine components $\ket{g_1}$ and $\ket{g_2}$ using the Stern-Gerlach effect. 

We then measured the ground state populations $N_1$ and $N_2$ via resonant absorption imaging following TOF, using our auxiliary imaging path [green lines in Fig.~\ref{Fig2:ExpSystem}(a)].
We quantitatively determine the atom number $N_1$ and $N_2$ in each state using bimodal fits to the TOF density distributions as described in Ref.~\onlinecite{Altuntas2022_Heating}, and denote the fraction in $\ket{g_2}$ as $f_2 = N_2/ (N_1 + N_2)$.
As a function of the relative phase shift $\delta\phi_P$ this fraction yields the characteristic cosinusoidal Ramsey oscillation from which we extract the light-induced phase shift $\phi$.

\subsection{Experimental imaging setups} \label{sec:ImagingSetups}

As illustrated in Fig.~\ref{Fig2:ExpSystem}(a) we employ two imaging configurations: a primary {\it in situ} configuration (red lines), and an auxiliary TOF configuration (green lines). 
Our technique calibrates the probe for the primary imaging configuration using information obtained through the auxiliary setup.
Both imaging paths utilize an initial Keplerian microscope (with magnification $\approx 9$), and the shared objective lens (L1) sets the numerical aperture to be 0.32.

The auxiliary path is separated using a small mirror placed near the focus of atoms after TOF, and concludes with a second ``microscope'' that reduces the magnification to 3.06(1).
This path utilized a PointGray Flea3 charge-coupled device (CCD)~\bibnote{Certain commercial equipment, instruments, or materials are identified in this paper in order to specify the experimental procedure adequately. 
Such identification is not intended to imply recommendation or endorsement by the National Institute of Standards and Technology, nor is it intended to imply that the materials or equipment identified are necessarily the best available for the purpose.}.
The primary imaging path uses a second stage microscope that sets a total magnification of about 36, and includes a phase dot positioned at the Fourier plane [as shown in Fig.~\ref{Fig2:ExpSystem}(a)] to enable phase-contrast imaging.
Our primary imaging sensor is an Andor DU-888UU3 electron multiplying CCD (EMCCD) with a $1024\times1024$ array of $13\ \mu{\rm m}$ square pixels.

Each CCD pixel converts incoming photons to photoelectrons with an efficiency given by the quantum efficiency ${\rm QE}$.
In this way $N$ photons yield $N_{\rm pe} = {\rm QE}\times N$ photoelectrons; in the case of an EMCCD, the resulting charge is amplified by electron multiplication.
In either case the resulting signal is digitized, and reported as $N_{\rm ADU}$ [in units of analog to digital units (ADUs)].
$N_{\rm ADU}$ has no particular relation to photo-electrons, nor photons and certainly not intensity at the location of the atoms.
In Sects.~\ref{sec:IntensityVary} and~\ref{sec:QEfactor}, we derive these relations. 

Even in the absence of direct illumination, our CCD accumulates photoelectrons; this results from a combination of background light and dark current.
To compensate this effect, for each measurement we acquire a dark image $N_{\rm ADU}^{{\rm dark}}$ with the beam off.
The dark image was subtracted from the bright-field images to remove dark counts and stray light\bibnote{All our measurements were conducted with the room lights off and the camera enclosed to minimize stray light.}.

\section{Ramsey interferometry measurements} 
\label{sec:RamseyLightShift}

\begin{figure}[tb!]
\begin{center}
\includegraphics{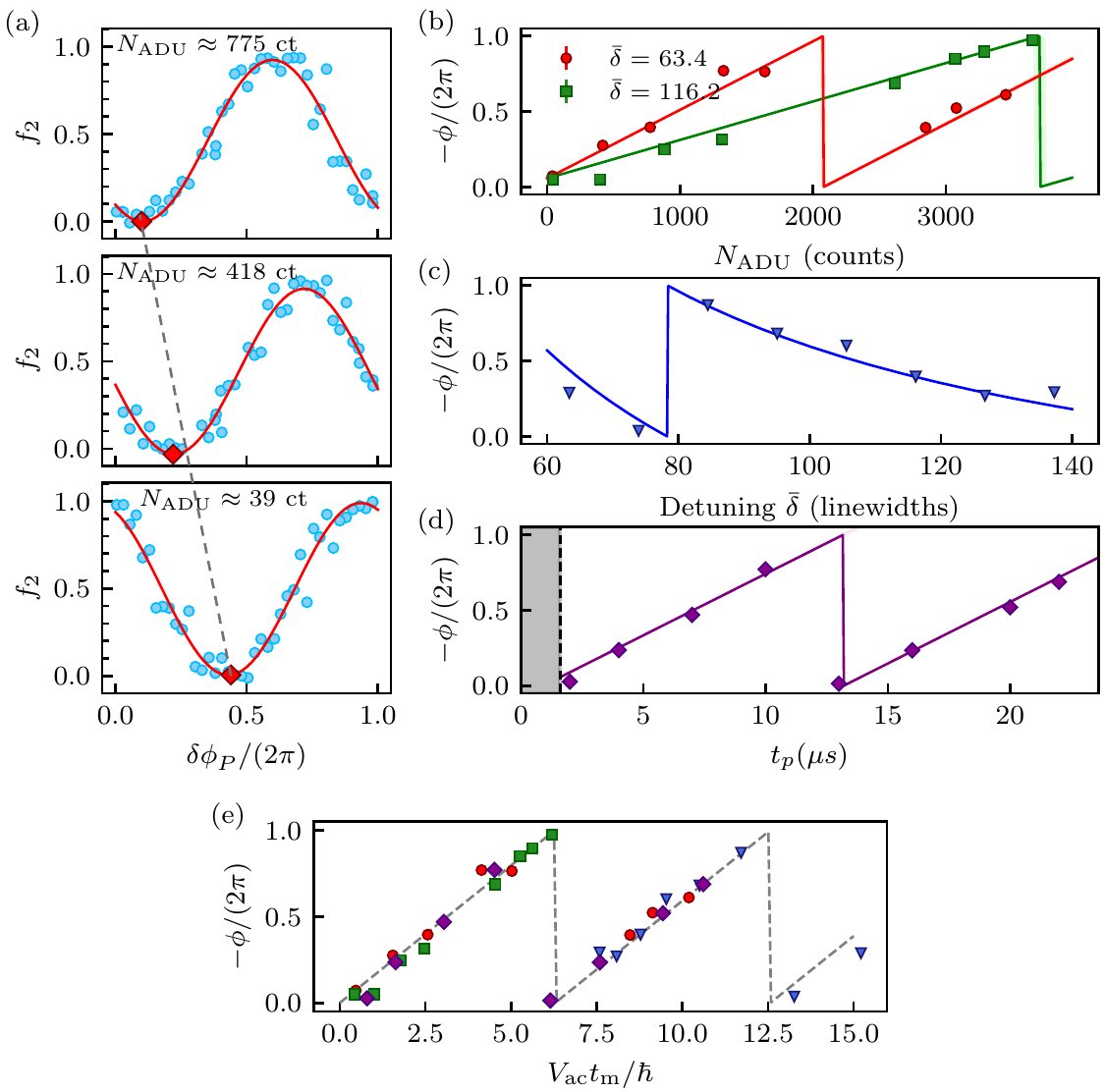}
\end{center}
\caption[Direct calibration of Nsat]{Ramsey interferometry.
(a) RI oscillations with probe power increasing from bottom to top, all at $\bar\delta = 63.4$ and $t_p = 20~\mu\rm{s}$. 
Red diamonds provide a guide for the optically-induced phase shift $\phi$.
(b) Phase shift $\phi$ as a function of probe intensity in units of $N_{\rm ADU}$, at $t_p = 20~\mu\rm{s}$ for $\bar\delta = 63.4$ (red) and $\bar\delta = 116.2$ (green).
(c) $\phi$ as a function of $\bar\delta$ at $t_p = 20~\mu\rm{s}$ and $N_{\rm ADU} = 5250(20)$.
(d) $\phi$ as a function of $t_p$ at $\bar\delta = 63.4$.
The probe intensity was set to give $N_{\rm ADU} = 3190(40)$ at $t_p = 20~\mu\rm{s}$.
The vertical dashed line marks the time $\delta t_0 = 1.6~\mu\rm{s}$ at which the probe pulse began.
(b-d) In all cases, same-color curves are fits to Eq.~\eqref{Eqn:PhaseShift_w12} mod $2\pi$.
(e) Data plotted as a function of $V_{\rm ac}t_{\rm m}/\hbar$ using the calibrated $N_{\rm sat}$ value.
The dashed gray line plots the expected outcome, a sawtooth curve: a line with slope $1$ mod $2\pi$.
Note that the vertical axes of (b)-(e) plot $-\phi$ for clarity.
}
\label{Fig3:RIL_Nsat}
\end{figure}

We now turn to the result of our RI measurements with data acquired as described in Sec.~\ref{sec:ExpSys}.
Figure~\ref{Fig3:RIL_Nsat}(a) plots measured RI oscillations (blue points) at three different beam powers with detuning $\bar\delta = \delta/\Gamma = 63.4$ and probe pulse time $t_p = 20~\mu\rm{s}$.
At low probe power [Fig.~\ref{Fig3:RIL_Nsat}(a), bottom] the phase shift is minimal, analogous to the black curve in Fig.~\ref{Fig1:Principle}(c).
The RI phase shift increases with increasing beam power [from middle to top in Fig.~\ref{Fig3:RIL_Nsat}(a)].
The resulting leftward shift of the curves is indicated by the red diamonds at the minimum of each oscillation.
The red curves depict fits to Eq.~\eqref{Eqn:RI_fringe_fit}, from which we determined the light-induced phase shift $\phi$.
 
Figure~\ref{Fig3:RIL_Nsat}(b) shows\bibnote{The data in Fig.~\ref{Fig3:RIL_Nsat}(b) were initially published in Ref.~\onlinecite{Altuntas2022_RIL}, all other data in this manuscript is previously unpublished.} $\phi$ obtained using this procedure as a function of $N_{\rm ADU}$ for two detunings $\bar\delta$.
These data exhibit $2\pi$ phase jumps because RI measures phase modulo $2\pi$.
We obtained $N_{\rm ADU}$ by imaging the probe beam without atoms on the primary imaging sensor for each RI sequence, and define $N_{\rm ADU}$ as the counts averaged within the sensor region where atoms would cast their shadow.
With respect to these units, $N_{\rm sat}$ is the number of ADUs resulting from a probe of intensity $\Isat$ illuminating the detector for $1\ \mu{\rm s}$.
In principle, $N_{\rm sat}$ can be obtained by fitting Eqs.~\eqref{Eqn:VAcStark} and \eqref{Eqn:RI_phase} to these data.
To increase the robustness of calibration, we added RI measurements with variable $\bar\delta$, constant $t_p$ and constant intensity [Fig.~\ref{Fig3:RIL_Nsat}(c)]; and variable $t_p$, constant $\bar\delta$, and constant intensity [Fig.~\ref{Fig3:RIL_Nsat}(d)].

In Eq.~\eqref{Eqn:RI_phase}, $t_m$ describes the true duration of the probe pulse. 
In our apparatus the control hardware introduces a dead time $\delta t_0$ [indicated by the vertical dashed line in Fig.~\ref{Fig3:RIL_Nsat}(d)] prior to applying the probe light.
This reduces the pulse time to $t_m = t_p - \delta t_0$.
We perform a joint fit to the full data set spanning Figs.~\ref{Fig3:RIL_Nsat}(b)-(d) yielding $N_{\rm {sat}} = 27.2(4)~\rm{counts/pix}/\mu\rm{s}$. 

In reality the model function is modified compared to Eq.~\eqref{Eqn:VAcStark} because $\ket{g_1}$ is also coupled to $\ket{e_2}$.
Although this transition is far-detuned compared to the cycling transition, its Stark shift cannot be neglected.
The total phase shift is 
\begin{align}
\phi = -\frac{\Gamma}{8} t_{\rm m} \frac{I}{\Isat} \left[ \frac{\Gamma}{\delta} - \frac{\Gamma}{2 \delta_{12}} \right],
\label{Eqn:PhaseShift_w12}
\end{align}
where $\delta$ is the probe detuning from the $\ket{g_2}$ to $\ket{e}$ transition, and $\delta_{12} = \delta - \Delta_G + \Delta_E$ is the detuning from the $\ket{g_1}$ to $\ket{e_2}$ transition, with ground state hyperfine splitting $\Delta_G \approx 6.8$ GHz and excited state hyperfine splitting $\Delta_E \approx 266\ {\rm MHz}$. 
This does not increase the number of free parameters.

In addition, these data are consistent with a small phase shift $\phi_0$ even for $V_{\rm ac} t_m/\hbar = 0$.
Experimentally we command $V_{\rm ac}=0$ or $t_m=0$ by setting an analog signal to an acousto-optic modulator (AOM) to zero; in practice this does not completely attenuate the AOM's radio-frequency drive, leading to a small amount of leakage light.
We confirmed that $\phi_0 = 0$ when the probe light is mechanically blocked.
As a result, we include $\phi_0$ as a free parameter in our joint fit, and added an auxiliary data set with $V_{\rm ac}=0$ consisting of 8 RI fringes to yield $\phi_0/(2\pi) = 0.06(2)$.

Fig.~\ref{Fig3:RIL_Nsat}(e) summarizes the result of our calibration by plotting the full data set as a function of $V_{\rm ac} t_{\rm m}/\hbar$. 
These data collapse onto a single saw-tooth curve. 
The dashed black curve in Fig.~\ref{Fig3:RIL_Nsat}(e) plots the expected phase shift $\phi$, a line with slope $1$ mod $2\pi$.
We observe good agreement between the data and the expected ac Stark shift scaling.
The two data points that deviate from the expected value were taken at our highest intensity and relatively close to resonance.

\section{Spatial variation of intensity} \label{sec:PixByPix}

In this section, we apply the RI techniques described in Sec.~\ref{sec:RamseyLightShift} on a pixel-by-pixel basis to characterize spatial variations in the probe intensity~\cite{Windpassinger_2008, Alberti2021}.
This analysis yields the Ramsey phase shift as a function of the average intensity, from which we deduce the fractional intensity difference throughout the atomic ensemble.  

\begin{figure}[htb]
\begin{center}
\includegraphics[width=5.2in]{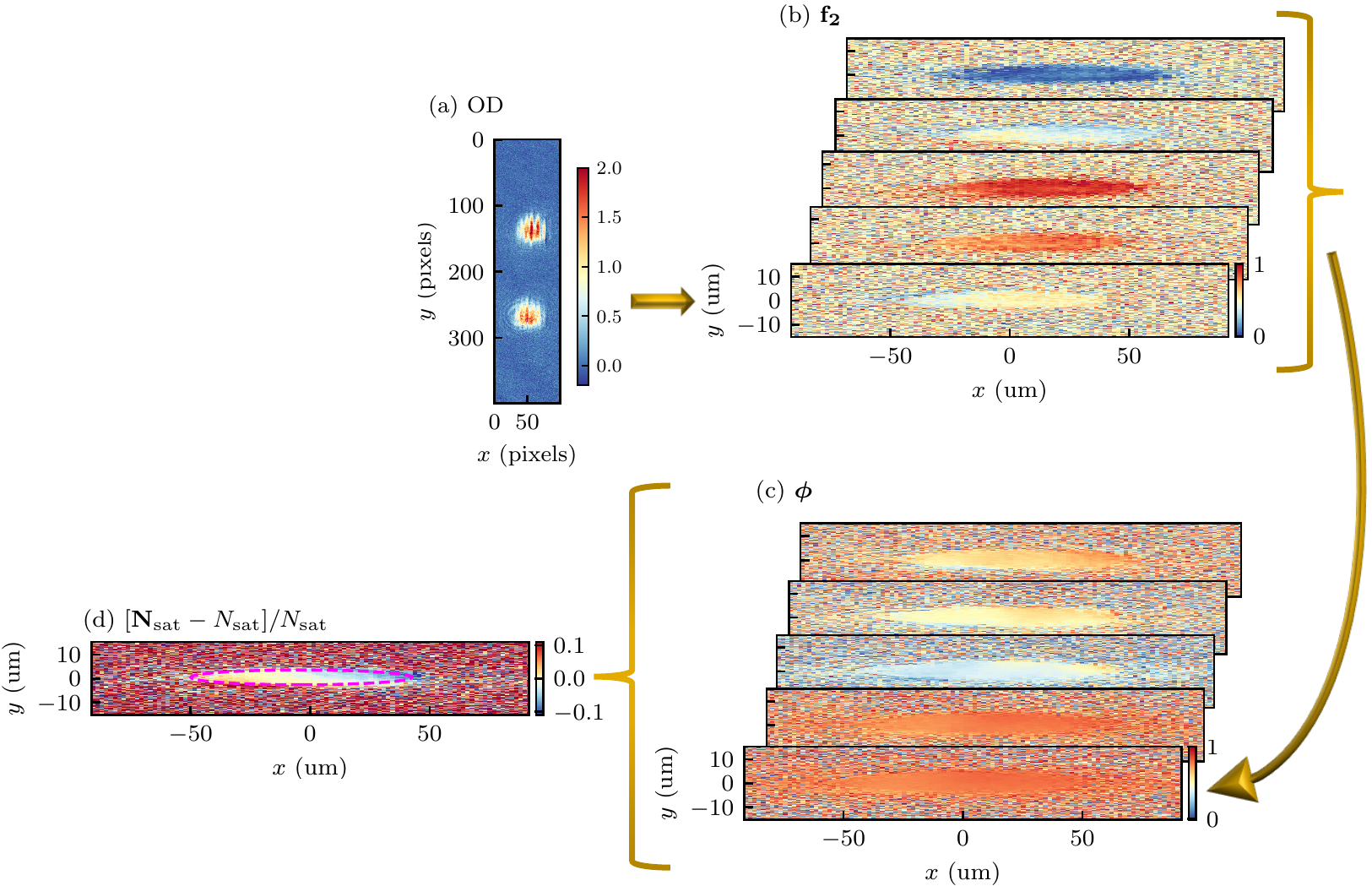}
\end{center}
\caption[Spatial variation of the fractional intensity]{Pixel-by-pixel probe calibration.
(a) Raw TOF image showing $\ket{g_1}$ and $\ket{g_2}$ clouds in TOF (in nonscaled pixel units), following the RI sequence.
In the remaining panels (b-d), the vertical axis is scaled from TOF to {\it in situ} coordinates as described in the text.
(b) ${\bf f_2}$ evaluated at five different values of $\delta \phi_P$ at $\bar\delta = 63.4$, $t_p = 20~\mu\rm{s}$ and averaged $N_{\rm {ADU}} = 1630(40)$.
The bottom image results from the data in (a).
(c) $\boldsymbol{\phi}$ at five values of $N_{\rm {ADU}}$, increasing from bottom to top.
The ensemble of ${\bf f_2}$ sampled in (b) leads to the bottom panel. 
(d) Fractional intensity difference derived from the set of $\boldsymbol{\phi}$ in (c). 
The dashed magenta ellipse signifies the region occupied by the BEC.
}
\label{Fig5:Nsat_pixBYpix}
\end{figure}

Figure~\ref{Fig5:Nsat_pixBYpix}(a) depicts a typical TOF image acquired in RI sequence with $f_2 \approx 0.5$. 
For each repetition of the experiment the overall center positions of these clouds varied on the $10\ \mu{\rm m}$ scale, although their relative positions did not.
Therefore to correctly associate pixels between the $\ket{g_1}$ and $\ket{g_2}$ clouds, we center each cloud within a rectangle region of interest prior to computing $f_2$ on a pixel-by-pixel basis; this gives an image ${\bf f}_2$.
Our centering process relied on fits to the 2D Thomas-Fermi (TF) profile to obtain center positions for both $\ket{g_1}$ and $\ket{g_2}$ clouds.

The magnetic field gradient used in the Stern-Gerlach separation of the atomic hyperfine components is created from a coil-pair in an anti-Helmholtz configuration.
As a result, the gradient field also induces a harmonic potential---which is trapping for $\ket{g_2}$ and anti-trapping for $\ket{g_1}$---during TOF.
This could potentially induce differential changes in the $\ket{g_1}$ and $\ket{g_2}$ cloud profiles.
Such an effect would cause the $\ket{g_1}$ cloud to be stretched along $\ex$ and $\ket{g_2}$ cloud to be compressed along $\ex$ (with no effect along $\ey$).
The widths from the 2D TF fits indicated that stretching effects were insignificant. 

During TOF, repulsive interactions cause the atomic clouds to expand predominately along $\ey$ (and $\ez$, which is not imaged) altering the aspect ratio.
We account for this effect by applying the Castin-Dum scaling theory~\cite{Castin1996} to TOF images, transforming from TOF to {\it in situ} coordinates.
In what follows the vertical axis is scaled, yielding the {\it in situ} distributions, for example the fractions ${\bf f}_2$ shown in Fig.~\ref{Fig5:Nsat_pixBYpix}(b).
In these data we minimize the noise contribution from pixels with no atoms by replacing data outside the range $f_2 \in [-0.1,1.1]$ with $f_2 = 0.5$ (in such pixels, $f_2$ is the ratio of two random numbers centered at zero and is essentially unbounded).

Furthermore, the raw data in (a) exhibit periodic density modulations.
This is a well known effect in quasi-1D BECs~\cite{Gerbier2003} and is not related to our intensity calibrations~\bibnote{These modulations are generically present: with and without the Ramsey pulse sequence; and with and without the {\it in situ} probe beam pulse.
This confirms that their origin is unrelated to inhomogeneities in the probe beam.}.
We remove these ``stripes'' by applying a Fourier space filter (a 2D elliptical Tukey window) to ${\bf f}_2$
that excludes only the specific frequency components associated with these stripes.
This ensures that the remaining structure in the data result only from the probe beam.
The displayed fractions in Fig.~\ref{Fig5:Nsat_pixBYpix}(b) were measured for five different values of $\delta \phi_P$ at $\bar\delta = 63.4$, and $t_p = 20~\mu\rm{s}$.
This is the same data presented in Fig.~\ref{Fig3:RIL_Nsat}(b) for which we obtained $N_{\rm {ADU}} = 1320(30)$ averaged over the whole cloud.

After this preliminary analysis, we determine the phase shift $\boldsymbol{\phi}$ by fitting $f_2(\delta \phi_P)$ to Eq.~\eqref{Eqn:RI_fringe_fit} for each pixel. 
Analogous to the procedure in Sec.~\ref{sec:RamseyLightShift}, this yields $\boldsymbol{\phi}$ versus the averaged $N_{\rm {ADU}}$; five examples are shown in Fig.~\ref{Fig5:Nsat_pixBYpix}(c).
In the final step, we fit the model function Eq.~\eqref{Eqn:PhaseShift_w12} on a pixel-by-pixel basis to $\boldsymbol{\phi}$, yielding the fractional intensity difference $[{\bf N}_{\rm {sat}} - N_{\rm {sat}}]/N_{\rm {sat}}$ shown in Fig.~\ref{Fig5:Nsat_pixBYpix}(d).
Within the region of the atomic ensemble (marked by the magenta dashed ellipse), our probe intensity is largest for negative $x$ and decreases for increasing $x$.

\section{Sensor calibration via shot noise} \label{sec:IntensityVary}

For any given camera, the relation between ADUs and photoelectrons is at best poorly documented but usually just unknown.
Here we connect these quantities by taking advantage of the ``shot noise'' of the photoelectrons whereby the variance of a signal with on average $N_{\rm pe}$ photoelectrons is also $N_{\rm pe}$, i.e., with uncertainty $\Delta N_{\rm pe} = \sqrt{N_{\rm pe}}$.
In addition, Ref.~\onlinecite{Robbins2003} showed that the stochastic EM process increases the noise $\Delta N_{\rm pe}$ by a factor of $\sqrt{2}$; we account for this by setting $\Delta N_{\rm pe} = \sqrt{2 N_{\rm pe}}$.

\begin{figure}[tb!]
\begin{center}
\includegraphics{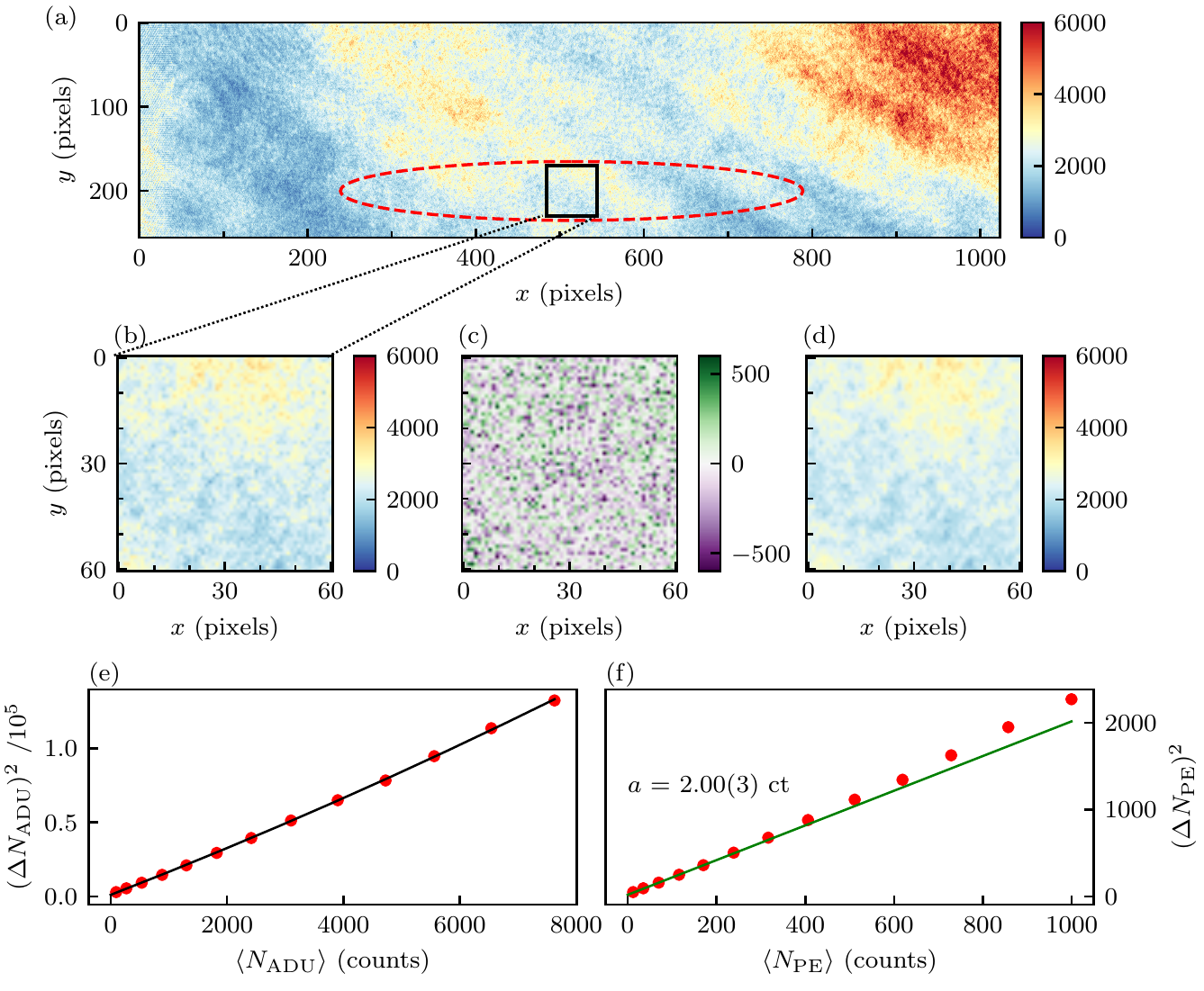}
\end{center}
\caption[Independent calibration of saturation intensity]{Sensor calibration via shot noise. 
(a) Raw probe image with inhomogeneities on many length scales. 
The red dashed ellipse signifies the typical location of the BEC's shadow (no atoms are present in these data).
The black square indicates the relatively homogeneous region of interest used in the noise analysis.
(b) Expanded view of the region of interest.
(c) Noise signal after subtracting the PCs from the probe image.
(d) Background signal: difference between (b) and (c).
(e) Variance and average intensity in units of ADUs. 
The black curve is the quadratic fit to the data.
(f) Variance and average intensity in units of photoelectrons computed using the conversion factor $C$.
The green curve, with slope $a = 2.00(3)$ is the linear contribution of a quadratic fit.
}
\label{Fig4:Nsat_Ivary_PCA}
\end{figure}

Therefore comparing the observed ADU signal $N_{\rm ADU}$ with its variance $\Delta N_{\rm ADU}^2$ allows us to obtain a direct conversion 
\begin{equation}
N_{\rm ADU} = C \times N_{\rm pe},
\label{Eqn:C_ADUtoPE}
\end{equation}
between $N_{\rm ADU}$ and $N_{\rm pe}$, with scale factor $C$. 
In principle, to extract $C$ we would simply illuminate the sensor with a uniform probe beam and measure the variance as a function of intensity.
In reality the probe beam pictured in Fig.~\ref{Fig4:Nsat_Ivary_PCA}(a) is inhomogeneous, with a fringe pattern that changes on the $\approx 1\ {\rm ms}$ time scale.
This necessitates the more complex analysis procedure described below.

We began by focusing our attention on a small, relatively homogeneous portion of the sensor [black square in Fig.~\ref{Fig4:Nsat_Ivary_PCA}(a)] within the region where our BEC casts its image (red dashed ellipse).
Then we use a principle component analysis (PCA) technique to distinguish between noise and overall changes in the probe profile. 
At each desired intensity we acquire an ensemble of probe images, and compute an orthonormal PCA basis.
The orthonormal components of the PCA basis are ordered by their relative contribution to the dataset; the first few components reflect structure in the probe and the remaining components reflect shot noise.

In our experiment, we obtained an ensemble of $n=35$ probe images $\{N_i\}_{i=1}^{35}$ (in ADU units) for a given intensity with average $\langle N \rangle$.
We then implement the following PCA based variance analysis procedure:
\begin{enumerate}  
\item For each element of this ensemble $N_i$, we construct a features matrix $X_i$ from the ensemble excluding $N_i$.
The features matrix has dimensions $(m,n-1)$, where $m$ is the sensor area in pixels.
\item We compute the covariance matrix from $X_i$, diagonalize the covariance matrix and obtain its eigenvectors and eigenvalues\bibnote{Given the large pixel count of our images, i.e., $m \gg n$, the common singular value decomposition algorithm was observed to be slow. To improve computational efficiency, we compute the covariance matrix as $X^T \cdot X$, where $T$ indicates transpose operation.}.  
These orthonormal ``principal'' components (PCs) are ordered by their relative contribution to the dataset (quantified by their eigenvalues).
This procedure yields a set of PCs, $\{{\rm PC}_{i,k}\}_{k=1}^{n-1}$ for each $N_i$.
\item We sum the contribution of each PC in $\{{\rm PC}_{i,k}\}_{k=1}^{n-1}$ to the image $N_i$ to obtain the ``averaged'' intensity $\langle N_{i}\rangle$.
By excluding $N_i$ from the features matrix, this average lacks the pattern of shot noise unique to this image.
Hence, this also gives the noise $\Delta N_{i} = N_i - \langle N_{i}\rangle$.
\end{enumerate}
Figures~\ref{Fig4:Nsat_Ivary_PCA}(c) and (d) depict the resulting noise and average intensity, respectively, from the raw data in (b).
We then repeat this procedure for a range of nominal intensities.

Figure~\ref{Fig4:Nsat_Ivary_PCA}(e) plots the variance as a function of the average ADU counts, $\langle N_{\rm ADU}\rangle$.
We expect this curve to be linear with a vertical intercept giving the sensor's read noise.
In addition, imperfections in the background removal process introduce a quadratic contribution.
We therefore fit this data to a quadratic function [black curve in (e)], and the linear term leads to the conversion factor $C = 7.65(2)$ for our primary CCD. 
At our largest intensity with $\langle N_{\rm ADU}\rangle \approx 8000$, the quadratic term is only a $12\ \%$ effect, and the read noise (in ADU) $\Delta N_{\rm read} = 32(19)$ is negligible.

Before developing our PCA based analysis, we implemented a simple-minded high-pass filter on the probe to remove the long-wavelength probe structure.
This straightforward approach lead to a significant quadratic component in the fits described above.
As such, the PCA-based algorithm was accepted as a superior tool for sensor calibration. 

Figure~\ref{Fig4:Nsat_Ivary_PCA}(f) cross-checks our calibration by plotting the variance and the mean in units of photoelectrons.
An ideal calibration would yield a line with a slope of $2$.
We cross-check these scaled data by fitting to the parabolic function described above and find $a = 2.00(3)$ as expected.
We plot only the linear contribution (green line); the deviation of this line from the data graphically indicates the scale of the quadratic contribution from imperfect background subtraction.

\section{Quantum efficiency and optical attenuation} \label{sec:QEfactor}

Manufacturers of scientific cameras provide QE curves as a function of wavelength, and in some cases at different sensor temperatures. 
However, large differences between the specified and measured QE values have been reported in the literature~\cite{Sperlich_2014}.
We conclude this study with a determination of the QE of our EMCCD.

The experimental procedure for QE factor measurement proceeds as follows.
We illuminate the EMCCD with a clean collimated Gaussian beam from an optical fiber and calibrate the beam power $P$ on a photodiode.
There are no optical elements between the fiber collimation package and the EMCCD sensor or the photodiode. 
We then image the beam on the EMCCD and relate the integrated $N_{\rm ADU}$ to the number of photons in a pulse of duration $t_p$.
We designed the beam to be appreciably smaller than the sensor, thus directly giving the conversion from photons to ADU.
Comparing this conversion to the established conversion between $N_{\rm ADU}$ and $N_{\rm pe}$ yields the QE.

\begin{figure}[tb!]
\begin{center}
\includegraphics{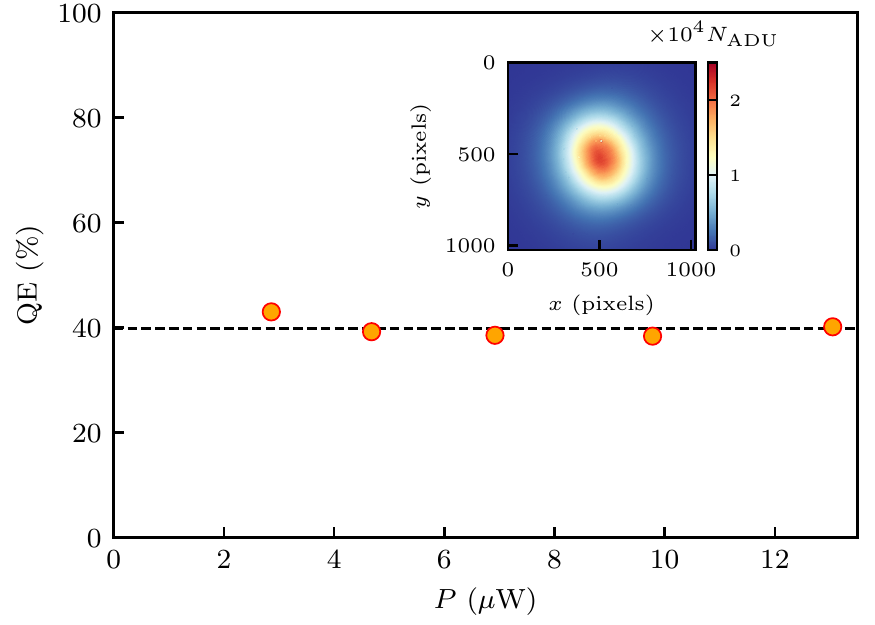}
\end{center}
\caption[]{QE factor of Andor DU-888UU3 with $-60\ {\rm C}$ sensor temperature.
Dashed black line indicates the average QE.
Inset: the near-perfect beam profile on the sensor, imaged with beam power $P=13.05(2)~\mu\rm{W}$.
}
\label{Fig6:QEfactor}
\end{figure}

The inset to Fig.~\ref{Fig6:QEfactor} shows a typical detected Gaussian beam. 
We limited the probe power to ensure that no pixels were saturated compromising the fidelity of the image.
We then fit a 2D Gaussian
\begin{align}
G(x,y) &= A_g \exp\left[-2\sum_{i=x,y}\left(\frac{x_i-b_i}{\sigma_i}\right)^2 \right]+d,
\end{align}
to the data, where $A_g$ is the amplitude, $\sigma_{x,y}$ are the widths, $b_{x,y}$ are center positions, and $d$ is a small offset.
The integrated $N_{\rm ADU} = \pi \sigma_x \sigma_y A_g /2$ is then determined from a direct integral of the fit.

The number of photons in the optical pulse
\begin{align}
N_{\rm ph} &= \frac{P t_m}{h c/ \lambda},
\end{align}
directly leads to the quantum efficiency ${\rm QE} = N_{\rm pe} / N_{\rm ph} = N_{\rm ADU} / (C \times N_{\rm ph})$.
For example, a power of $13.05(2)~\mu\rm{W}$ leads to $N_{\rm ph} = 9.5\times 10^8$ and $N_{\rm ADU} = 2.9\times 10^9$, giving ${\rm QE} = 40.1(8) \% $.
Figure~\ref{Fig6:QEfactor} shows that the QE measured in this way is independent of the $P$, thereby further validating the method. The variation in QE seen in Fig.~\ref{Fig6:QEfactor} of $2.9\ \%$ is consistent with our statistical uncertainty estimate and leads to the average value ${\rm QE} = 40(2)$. 

This measured QE is consistent with an independently measured value for our EMCCD~\cite{Sperlich_2014}.
This value is only $61\ \%$ of manufacturer's claimed QE for our operating temperature $-60 ^\circ {\rm C}$. 

\subsection{Optical attenuation}

In addition to the effective loss from the QE (and the noise added from the EM gain stage), the many optical elements in our imaging system [Fig.~\ref{Fig2:ExpSystem}(a)] attenuate the probe.
Our measurement of $N_{\rm sat}$ along with the conversion from $N_{\rm ADU}$ to $N_{\rm pe}$ allows us to obtain the transfer efficiency $T$ of our imaging system.

The relationship between $N_{\rm ADU}$ and intensity at the object plane is
\begin{equation}
I = \left[\frac{1}{C \times {\rm QE}\times T} \frac{1}{A \times t_m}\frac{hc}{\lambda}\right]N_{\rm ADU},
\label{Eqn:I_to_Nadu}
\end{equation}
where $A = (13\ \mu{\rm m} / M)^2$ is the area of a single CCD pixel (as demagnified to the object plane); $t_m$ is the pulse time; and $h$ is Planck's constant.
This expression can be inverted to give the total system efficiency 
\begin{align}
{\rm QE}\times T &= \left[\frac{1}{C}\frac{1}{A \times 1\ \mu {\rm s}} \frac{hc}{\lambda} \right]\frac{N_{\rm sat}}{\Isat} \lesssim 0.4
\end{align}
where we used $t_m = 1\ \mu {\rm s}$ in accordance with our definition of $N_{\rm sat}$.

\section{Conclusion and outlook}
\label{sec:Conc}

Here, we presented a new technique for determining the saturation intensity for cold-atom experiments and explored its applications limitations.  
These measurements show that even a carefully designed imaging system can have significant losses; in our case the full system efficiency is below $40\ \%$.
For many applications, such as absorption imaging, this can be acceptable, however, in experiments targeting quantum back-action limited weak measurements, high system efficiency can be a priority.
For example, in quantum gas experiments, losses---as well as the added EM noise---have deleterious effects on dispersive weak measurement protocols where detected photon shot noise correlates with quantum backaction on the atomic system.

Common to all existing methods determining $\Isat$, our technique is sensitive to the polarization of the probe.
For example, in our case the light shift is maximized for $\sigma_+$ polarization.
We estimate our polarization purity---optimized by maximizing both absorption and phase contrast imaging signals---to be $\gtrsim 95\ \%$.
Indeed polarization effects can be quite significant because the ratio of the matrix elements squared for $\sigma_+$ versus $\sigma_-$ transitions is 15, meaning for example $I_{\rm sat} \approx 25.5\ {\rm mW}/{\rm cm}^2$ for $\sigma_-$ polarization. 
Thus a fractional polarization error $\epsilon$ scales the effective saturation intensity  by $(1-14 \epsilon /15)^{-1}$.
This polarization dependence prevents our method from being a primary measure of intensity, however, extensions to the method that measure the light shift for different orientations of the magnetic field could overcome this limit.

\begin{backmatter}
\bmsection{Funding}
This work was partially supported by the National Institute of Standards and Technology, the National Science Foundation through the Quantum Leap Challenge Institute for Robust Quantum Simulation (grant OMA-2120757), and by the Air Force Office of Scientific Research's Multidisciplinary University Research Initiative ``RAPSYDY in Q'' (grant FA9550-22-1-0339).

\bmsection{Acknowledgments}
The authors thank Alan Migdall and Daniel Barker for carefully reading the manuscript.

\bmsection{Disclosures}
The authors declare no conflicts of interest.

\bmsection{Data availability} 
Data underlying the results presented in this paper are not publicly available at this time, but may be obtained from the authors upon reasonable request.

\appendix

\section{Signal-to-noise ratio}
\label{App:SNR}

In this appendix, we derive the signal-to-noise ratio for resonant absorption imaging. 
The absorption of resonant laser light propagating along ${\bf e}_z$ traversing an atomic cloud with 3D number density $\rho$ derives from
\begin{align}
\frac{\mathrm{d}I(z)}{\mathrm{d}z} &= -\sigma_0 \rho(z) \frac{I(z)}{1+I(z)/\Isat},
\label{Eqn:Beer_Isat}
\end{align}
where $\sigma_{0} = 3 \lambda^2 /(2 \pi)$ is the resonant scattering cross-section expressed in terms of the laser wavelength $\lambda$.
For $I_{\rm sat}\rightarrow\infty$ the above expression reduces to the Beer-Lambert law.
Integrating Eq.~\eqref{Eqn:Beer_Isat} along $z$, yields the 2D column density 
\begin{align}
\sigma_0\rho_{\rm 2D}  = \sigma_0 \int \rho(z) \mathrm{d}z = -{\rm ln}\left(\frac{I_+}{I_-}\right) - \frac{I_+ - I_-}{\Isat} \equiv {\rm OD}_{\rm corr},
\label{Eqn:OD_Isat}
\end{align}
where $I_{-}$ is the intensity just prior to the atomic ensemble and $I_+$ is the intensity just following the atomic ensemble.
The dimensionless column density is given by the optical depth ${\rm OD} = -{\rm ln}\left(I_{+} / I_{-}\right)$ combined with a correction term accounting for $I_{\rm sat}$.
For convenience we define the combination to be a corrected optical depth ${\rm OD}_{\rm corr}$.

As detailed in Sec.~\ref{sec:ImagingSetups}, in our experiments we measure the probe laser's intensity on a CCD or an EMCCD.  
In each experimental shot we acquire three images: (1) $N_{\rm A}$, the probe field with the atoms present; (2) $N_{\rm P}$, the probe without the atoms; and (3) $N_{\rm D}$, a dark image with the probe beam off.
Each of these are taken to be in units of photoelectron counts.
In order to eliminate any baseline from stray light and CCD dark counts, we subtract $N_{\rm D}$ from $N_{\rm A}$ and $N_{\rm P}$ to obtain 
\begin{align}
N_{\pm} &=  \left({{\rm QE}\times T}\times{A \times t_m}\frac{\lambda}{hc}\right) I_\pm,
\end{align}
having used Eq.~\eqref{Eqn:I_to_Nadu} (scale factor $C$ omitted) to relate photoelectron counts to intensity.

As noted in Sec.~\ref{sec:Introduction}, the prevalent source of noise in bright-field measurements is the shot noise of the photoelectrons. 
Accordingly, we describe each photoelectron counts measurement $N = \langle N\rangle+\delta N$ as the sum of an ensemble average $\langle N\rangle$ and measurement noise $\delta N$.
We model $\delta N$ as a classical random process with zero mean $\langle\delta N \rangle = 0$ and variance $\langle\delta N^2 \rangle = \langle N \rangle$.

In experimental practice, we determine $N_-$ (from the probe beam without the atoms) by averaging over an ensemble of such measurements, making its noise contribution negligible.
Then, following Eq.~\eqref{Eqn:OD_Isat} we obtain 
\begin{align}
{\rm OD}_{\rm corr} = -{\rm ln}\left(\frac{N_{+} + \delta N_{+}}{N_{-}}\right)  - \frac{N_{+} + \delta N_{+} - N_{-}}{N_{\rm sat}}.
\label{Eqn:OD_withNoise}
\end{align}
Defining the ideal optical density without noise as ${\rm OD}_{\rm corr}^{(0)}$, allows us to recast this expression
\begin{align}
{\rm OD}_{\rm corr} = {\rm OD}_{\rm corr}^{(0)} -{\rm ln}\left(1 + \frac{\delta N_{+}}{N_{+}}\right)  - \frac{\delta N_{+}}{N_{\rm sat}}.
\end{align}
We assume that the noise term is small compared to $N_{+}$, and Taylor expand the second term to first order obtaining 
\begin{align}
{\rm OD}_{\rm corr} = {\rm OD}_{\rm corr}^{(0)} - \delta N_{+}\left( \frac{1}{N_{+}} + \frac{1}{N_{\rm sat}} \right),
\end{align}
where noise contributions are isolated from the ideal optical density signal. 
The noise in the detected optical density is then
\begin{align}
\delta{\rm OD}_{\rm corr} = \frac{1}{\sqrt{N_+}}  \left( 1 + \frac{N_+}{N_{\rm sat}} \right).
\label{Eqn:OD_noise}
\end{align}

Equation~\eqref{Eqn:OD_withNoise} can be simplified to
\begin{align}
{\rm OD}_{\rm corr} &\approx - \ln\left(\frac{N_{+}}{N_{-}}\right), & {\rm and} && {\rm OD}_{\rm corr} &\approx \frac{N_-}{N_{\rm sat}}\left(1 - \frac{N_{+}}{N_{-}}\right)
\end{align}
in the limits of $N_-\ll N_{\rm sat}$ and $N_-\gg N_{\rm sat}$ respectively.
In addition, the latter limit implies that $N_+ / N_- \approx 1$. 
From these expressions, we arrive at the simple results
\begin{align}
{\rm SNR} &\approx \sqrt{N_-} e^{-{\rm OD}_{\rm corr}/2} {\rm OD}_{\rm corr}, & {\rm and} && {\rm SNR} &\approx\frac{N_{\rm sat}}{\sqrt{N_-}} {\rm OD}_{\rm corr}
\end{align}
for the SNR in the respective limits.
These expressions enunciate the extreme limits considered in the main text: (1) for $I \ll \Isat$ noise is proportional to $1/\sqrt{I}$ and the SNR scales like $\sqrt{I}$ and (2) for $I \gg \Isat$ noise is proportional to $\sqrt{I}$ and the SNR scales like $1/\sqrt{I}$. 
On the other hand, in the regime $I \approx \Isat$, noise has the minimum possible value with SNR reaching to its maximum.

\section{Microwave pulse phase shift}
\label{App:UwavePhaseShift}

This appendix outlines our microwave control setup and how the $\delta\phi_P$ phase shift is implemented. 
We generate the microwave carrier using a Stanford Research Systems SG384 signal generator.
Next, this signal is mixed with an $\approx 100~\rm{ MHz}$ signal from a Novatech 409B direct digital synthesizer using a Marki IRW0618 single sideband mixer, enabling dynamical control of the microwave frequency.
The signal's amplitude is then controlled with a voltage-controlled attenuator (General Microwave Herley D1956);  a high power microwave amplifier (Microwave Amplifiers AM53) increases the amplitude to enable $\approx 10\ {\rm kHz}$ Rabi frequencies.
The high power signal passes though a microwave circulator-isolator (MCLI CS-57) that prevents back reflections from damaging the amplifier. 
Any reflected signal is diverted to the circulator's rejection port where it is attenuated and monitored with a power detector (Minicircuits 42 ZX47-40-S+). 
Following the circulator, the main signal is impedance matched with a stub tuner (Maury Microwave 1819C) and delivered to the atoms using the microwave horn illustrated in Fig.~\ref{Fig2:ExpSystem}(a).

\begin{figure}[tb!]
\begin{center}
\includegraphics{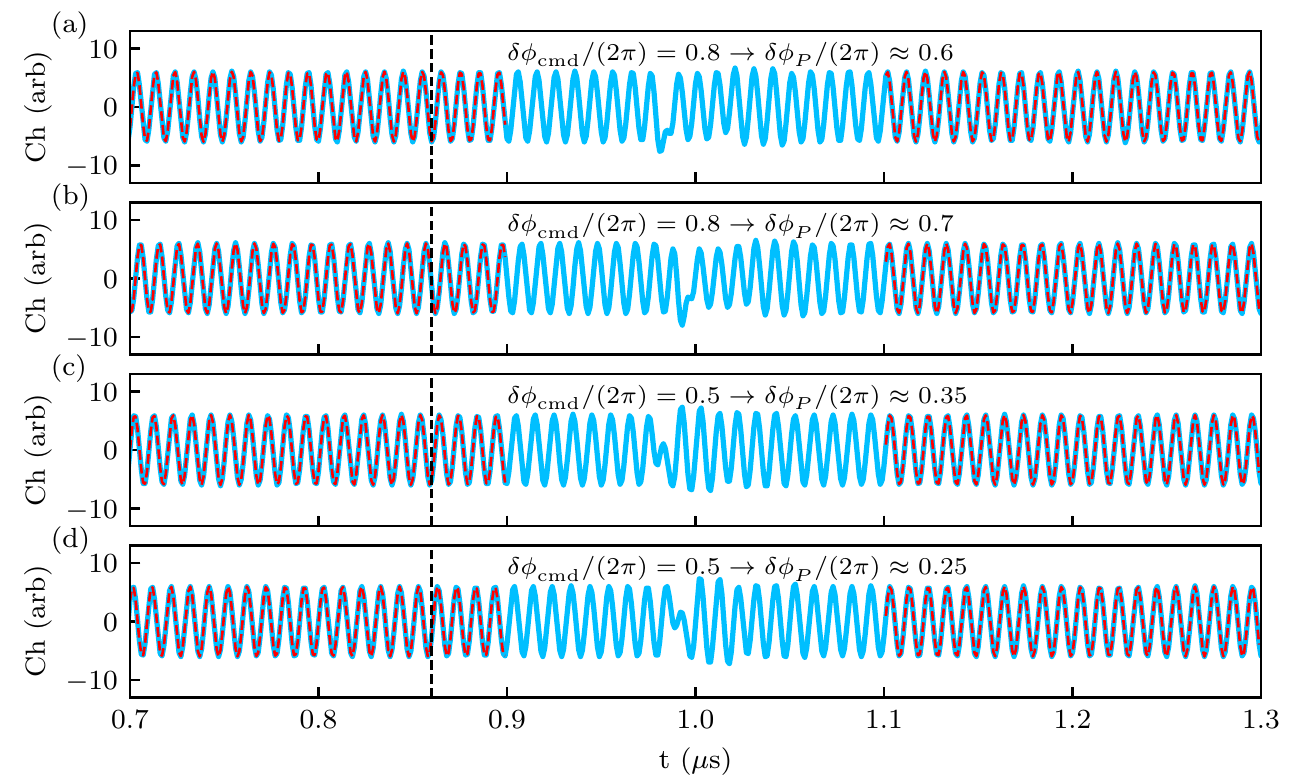}
\end{center}
\caption[Phase shift determination from oscilloscope measurements]{
Phase difference determination.
Blue curves are raw oscilloscope traces.
The red dashed curves fit Eq.~\eqref{Eqn:ScopeSine} to the signal before and well-after the phase update. 
The vertical dashed line marks the time at which the first $\pi/2$ microwave pulse ends.
(a,b) commanded phase shift of $\delta\phi_{\rm cmd} / (2\pi) = 0.8$.  
(c,d) commanded phase shift of $\delta\phi_{\rm cmd} / (2\pi) = 0.5$.
}
\label{FigApp:Scope}
\end{figure}

RI depends on the phase difference between the two microwave pulses (but not the absolute phase of the first pulse).
We implement this requirement by commanding a radio frequency (RF) phase of zero prior to the first microwave pulse, and then commanding $\phi_{\rm cmd}$ to the Novatech 409B between the two microwave pulses.
The astute reader might believe that $\delta\phi_P = \phi_{\rm cmd}$, however, hardware limitations thwart this expectation.

We employ the synchronous aligned phase update mode of Novatech to control the timing of the phase update at $\approx 10\ {\rm ns}$ level.  
In this mode, the oscillation waveform is $\propto \sin[2\pi f (t - t_0) + \phi_{\rm cmd}]$, where $t_0$ is the time at which the Novatech updates.
As such jitter in the update time manifests as unwanted phase shifts.
Figure~\ref{FigApp:Scope} shows four RF time traces with this jitter present.

Rather than upgrading the hardware to overcome this jitter, we directly measure the phase shift $\delta\phi_p$ for every experimental shot by fitting sinusoids
\begin{align}
g(t) = A\sin(2\pi f t + \pi \phi_e) + g_0,
\label{Eqn:ScopeSine}
\end{align}
to the RF waveform just before and just after the novatech update (red dashed curves).
Here $A$ is the amplitude, $f$ is the Novatech frequency, $\phi_e$ is the phase, and $g_0$ is an offset.
Since $f$ is known, the fit constraints $f$ value be within $0.002~\rm{MHz}$ of this value (we allow $f$ to vary to obtain an uncertainty estimate).
We thereby extract $\phi_e^{(\pm)}$ and find their difference $\delta\phi_P = \phi_e^{(+)} - \phi_e^{(-)}$.

The scope traces in Fig.~\ref{FigApp:Scope} illustrate two types of timing defects. 
First Fig.~\ref{FigApp:Scope}(a) and (b) show that the RF phase prior to the update pulse can be different from shot-to-shot (resulting from up-stream timing differences in different experiments).
Indeed the scope-fits confirm that the true phase differences differ from what was commanded to the Novatech.
Secondly, Fig.~\ref{FigApp:Scope}(c) and (d) show that even for equal initial phases, the observed phase difference can differ from the command.  
This results from the trigger timing jitter apparent in the figure. 
Notwithstanding, the fit procedure yields the exact phase shift as is required for high quality RI measurements.
\end{backmatter}

\bibliography{main}

\end{document}